%% file: main.tex
\documentclass[conference]{IEEEtran}
\IEEEoverridecommandlockouts
\usepackage{academicons}
\usepackage{cite}
\usepackage{graphicx}
\usepackage{comment}
\usepackage{balance}
\usepackage{acronym}
\usepackage{amsmath}
\usepackage{tabularx}
\usepackage{multirow}
\usepackage{amssymb}
\usepackage[usenames,dvipsnames]{xcolor}
\usepackage{subcaption}
\usepackage{booktabs}
\def\BibTeX{{\rm B\kern-.05em{\sc i\kern-.025em b}\kern-.08em
    T\kern-.1667em\lower.7ex\hbox{E}\kern-.125emX}}
    
\title{SingIt! Singer Voice Transformation}

\author{\IEEEauthorblockN{Amit Eliav}
\IEEEauthorblockA{\textit{Faculty of Engineering} \\
\textit{Bar-Ilan University}\\
Ramat-Gam, Israel \\
amiteli@biu.ac.il}
\and
\IEEEauthorblockN{Aaron Taub}
\IEEEauthorblockA{\textit{Faculty of Engineering} \\
\textit{Bar-Ilan University}\\
Ramat-Gam, Israel \\
aarontaub92@gmail.com
}
\and
\IEEEauthorblockN{Renana Opochinsky}
\IEEEauthorblockA{\textit{Faculty of Engineering} \\
\textit{Bar-Ilan University}\\
Ramat-Gam, Israel \\
renana.klainman@biu.ac.il
}
\and
\IEEEauthorblockN{Sharon Gannot}
\IEEEauthorblockA{\textit{Faculty of Engineering} \\
\textit{Bar-Ilan University}\\
Ramat-Gam, Israel \\
sharon.gannot@biu.ac.il
}

}

\input{definitions}
\input{acronyms}

\begin{document}

%\ninept

\maketitle

\begin{abstract}
\label{abstract}
In this paper, we propose a model which can generate a singing voice from normal speech utterance by harnessing zero-shot, many-to-many style transfer learning. Our goal is to give anyone the opportunity to sing any song in a timely manner. We present a  system comprising several available blocks, as well as a modified auto-encoder, and show how this highly-complex challenge can be achieved by tailoring rather simple solutions together. We demonstrate the applicability of the proposed system using  a group of 25 non-expert listeners. Samples of the data generated from our model are provided. 
\end{abstract}

\section{Introduction}
 With the latest explosion in the field of generative neural networks, this paper focuses on the ``musical capabilities'' of  deep networks. Machine learning algorithms have already impacted the world of music, mainly in the form of musical arrangements, instrumental backings, sound effects, and sound processing \cite{sturm2019machine, fiebrink2016machine}. Here we address the task of speech-to-singing transfer. We aim at a system that can enable us to hear what it would sound like if any given person was to sing any given song, or at least
 could be fun for those who have no musical talent at all, yet have a strong passion for singing.
%  To provide clarity, in this paper, we will utilize the terms Person A and Person B. Person A refers to the singer in the audio file, where the input consists of singer Person A's vocals along with the instrumental accompaniment. On the other hand, Person B refers to the \emph{user} whose input is solely speech, unrelated to the content of Person A's song. The output of our system is the content of the song sung by Person A but in the style of Person B. In other words, it generates the impression of Person A's song as if it were sung by Person B. 
%   More practically, we aim to apply the style transfer based only on the user's speech. This in itself has many interesting applications, in movies and other media, and more.
% %
\begin{figure}[htbp]
    \centering    \includegraphics[width=0.95\columnwidth]{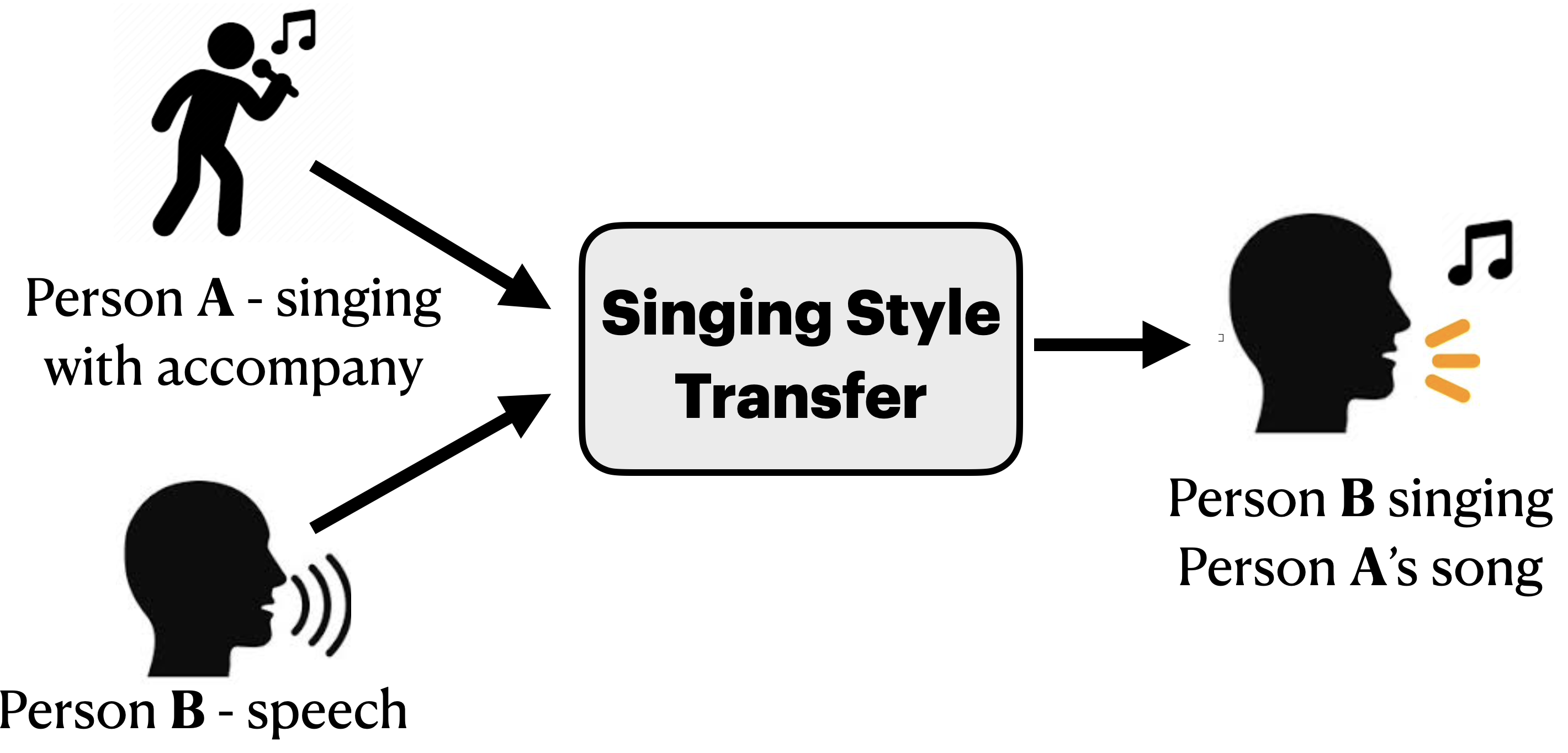}
    \caption{High level system overview.}
    \label{fig:High Level Overview}
\end{figure}
There are currently some other existing solutions attempting to solve similar problems, one of them being the \ac{NPSS} \cite{blaauw2017neural}, which has trained a model to learn how a specific singer sounds, and given a new song it produces the input song performed by the singer the network has trained on. There are other existing models for singing synthesis \cite{zhuang2021litesing} and singing from speech such as WGAN-Sing \cite{WGAN-Sing} which presents a deep neural network-based singing voice synthesizer. This network trains on a set of singers and has the ability to exchange styles and content but only within the singers in the database. Finally, AutoVC: Zero-shot voice style transfer \cite{autovc} uses an autoencoder architecture model to transfer between the voice signals of two speakers. Although this paper does not mention the use of the network for singing or music, we found that the architectural solution presented in this paper for style conversion of speech is very suitable for the approach we have chosen to adopt, addressing the problem of singing voice transfer. Our work can be regarded as an extension of the work in \cite{autovc} to the speech-to-singing task. To our best knowledge, such an extension cannot be found in the literature.

Our task and a schematic view of the system can be seen in Fig.~\ref{fig:High Level Overview}, where the two input voices, the style transfer block, and the output signal are indicated. 
Since we aim at a system that can be used in practice, we should be able to transfer the 
 voice of an unknown person. Moreover, the system should execute the style transfer in a timely manner.
 % The core of our system is an autoencoder that is fed by both the speaking voice (Person B) and the singing voice (Person A) and uses their embeddings to execute the style transfer. Subsequently, a vocoder network generates a waveform of Person B singing Person A's song. 
 % %
% In general, our system combines four main blocks: 1) A vocal separator - which splits the singing vocals from the instrumental accompaniment, 2) an embedding block, which outputs a vector representing the style of a given audio input, 3) an autoencoder which performs the style transfer and 4) a vocoder to convert the output spectrogram (of the already performed style transfer) to a waveform.
% We trained our network on a dataset that consists of both speech and singing audio samples in order for the network to become familiar with both types of audio. 
%

% Our model has the capability to convert any given speaker, to sing any given song, in a minimum amount of time. These requirements forced us to complexify the model architecture, as well as precisely plan the dataset and train stage for this model.  

 \section{Problem Statement}
 We will now rephrase the addressed problem in more specific terms. In this paper, refer to Person A and Person B. Person A refers to the singer in the audio file, where the input consists of singer Person A's vocals along with the instrumental accompaniment. On the other hand, Person B refers to the \emph{user} whose input is solely speech, unrelated to the content of Person A's song. The output of our system is the content of the song sung by Person A but in the style of Person B. In other words, it generates the impression of Person A's song as if it were sung by Person B. 
  More practically, we aim to apply the style transfer based only on the user's speech. This in itself has many interesting applications, in movies and other media, and more.
\section{Our Model}
\label{sec:Chosen_Architecture}
To address the challenge of converting an individual's style to a given song using only examples of speech, we had to deal with multiple sub-problems: 1) Vocal extraction: extracting the vocal channel from a song mix, 2) Speaker representation: taking one’s voice and describing it in a low-dimensional form, 3) Content embedding: Taking a song to a lower dimension where the singer's identity is no longer present, and 4) Style transfer: in the song’s lower dimension representation, switching between the singer’s style and the speaker’s style.
We tried to solve each sub-problem separately based on existing solutions and then combine them into one system.

\begin{figure*}[htbp]
    \centering      \includegraphics[width=0.84\textwidth]{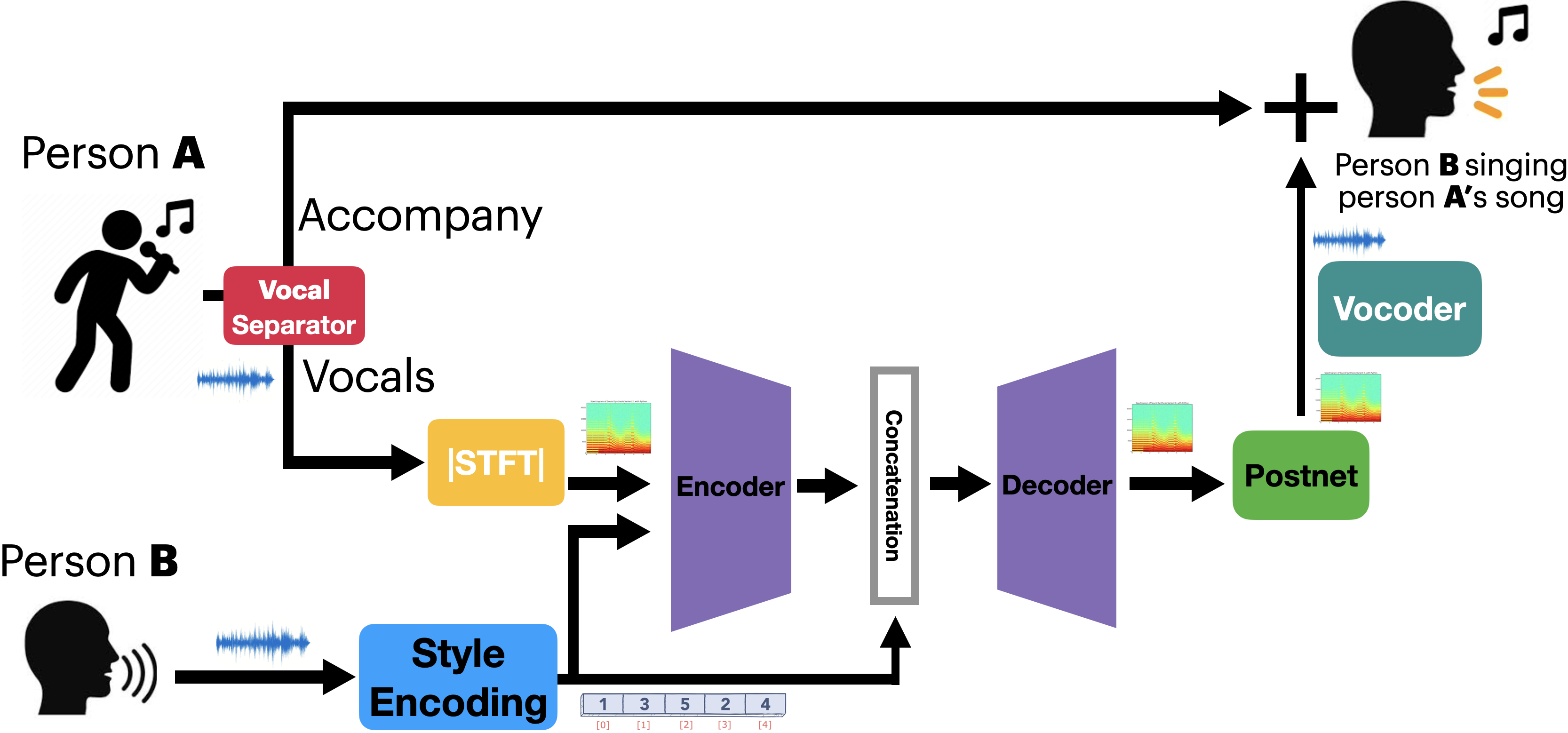}
      \caption{Detailed Solution Architecture.}
      \label{fig:General Solution Architecture}
   \end{figure*}

\sloppy The model is based on an autoencoder with self-reconstruction losses. That means that the model is not restricted to using parallel data which has both the speaker’s speech and the speaker’s singing and allows us to use existing datasets. A detailed description of the proposed system is depicted in Fig.~\ref{fig:General Solution Architecture}.
We will now elaborate on each of these building blocks.

%\begin{figure}[h]
%    \centering
%    \includegraphics[width=0.9\columnwidth]{General_Solution_Architecture.png}
%    \caption{General Solution Architecture}
%    \label{fig:General Solution Architecture}
%\end{figure}
%
\subsection{System Components}
\smallskip\noindent\textbf{Vocal extraction:}
\label{sec:Vocal’s Extraction}
 For the vocal separator, we used ‘Spleeter’ \cite{Spleeter, ca2021research}, a Python package that receives a song (mix of vocals and instruments) as an input and can separate the song into different stems (channels). In our case, we are separating each song into two stems: ‘vocal’ and ‘instrumental’.
%\vspace{-5pt}

\smallskip\noindent\textbf{Speaker representation:}
\label{sec:Speaker representation’}
We chose to use ‘Resemblyzer’,\footnote{\texttt{https://github.com/resemble-ai/Resemblyzer}} \cite{8462665}
% {$https://github.com/resemble-ai/Resemblyzer$}} 
a deep learning network that derives a high-level representation of a voice in the form of a 256-sized embedding vector. It should represent only the identity/style of the speaker regardless of the speech content and should summarize the characteristics of the given voice.

\smallskip\noindent\textbf{Content embedding:}
\label{sec:Content embedding}
We chose to use an encoder block with an architecture that follows the encoder block from \cite{autovc}. The content embedding model is a key part of our system, the bottleneck size and the input size were tested with many different values and the results were significantly different from one to another.

The \ac{STFT} log-spectrogram was preferred rather than the mel-spectrogram which is commonly used in many other speaker-to-speaker methods presented by other researchers. The reason is that the mel-scale tends to emphasize the speech-related frequencies, but in our case, singing voice, the log-spectrogram performed better.

The input to the encoder is a concatenation of the following two: 1) A log-spectrogram of Person A's song, with a shape of \(256\times T\), which are the number of \ac{TF} bins in the spectrogram, and 2) A 256-sized embedding vector, representing the target style. This embedding vector is replicated and concatenated to each time bin of the spectrogram. This result with an input of the size \(512\times T\).

Following \cite{autovc}, the encoder itself consists of 3 sets of a 1-D convolution layer, a batch normalization, and a Relu activation function, which is then followed by a \ac{BLSTM} layer. The output of these layers is then down-sampled by a factor of 32 and considered as the encoder's bottleneck. We experimented with the downsampling factor to achieve the best results.

%\vspace{-5pt}

\smallskip\noindent\textbf{Style transfer:}
\label{sec:Style transfer}
The style transfer is carried out in the decoder block, which also follows the decoder block architecture presented in \cite{autovc}. 
The decoder is responsible for reversing the process of the encoder. 

The input to the decoder is a concatenation of two inputs, namely the encoder output, and the 256-sized embedding vector, representing the target style.
This input passes a \ac{BLSTM} layer followed by 3 blocks, as in the encoder, namely a 1-D convolution layer, a batch normalization, and a ReLu activation function. Finally, we apply a second \ac{BLSTM} layer and a linear projection layer, which results in a spectrogram with the same \ac{TF} output size as the encoder's input spectrogram.
This output is a converted spectrogram. To complete the process, all that is needed is to convert the spectrogram back to a waveform, as explained in Sec.~\ref{sec:Vocoder}.

\smallskip\noindent\textbf{Postnet}
\label{sec:Postnet 1}
 Similar to the model described in \cite{autovc, 8461368}, we also apply another \ac{CNN}-based model to further enhance the quality of the spectrogram generated by the decoder.

\subsection{Loss Functions}
\label{sec:Losses}
We denote the log-spectrogram input as $X$, the encoder, decoder, and Postnet as 
\( \textrm{E}(\cdot), \textrm{D}(\cdot), \textrm{P}(\cdot) \), respectively. We also define $\textrm{codes}=\textrm{E}(X)$, $\hat{X}=\textrm{D}(\textrm{codes})$, and $\tilde{X}=\textrm{P}(\hat{X})$.
Three loss functions were used to train the model. %

\noindent \textbf{Loss 1:} An MSE loss calculated between $X$ and the decoder output $\hat{X}$:
    \begin{equation}
       L_1=\textrm{MSE}(X,\hat{X})
    \end{equation}
\textbf{Loss 2:} An MSE loss, calculated between $X$ and the Postnet model output $\tilde{X}$:
    \begin{equation}
       L_2=\textrm{MSE}(X,\tilde{X})
    \end{equation} 
\textbf{Loss 3:} An $\ell_1$ loss, calculated between the output of the encoder with the input data and the output of the encoder with the input taken from Postnet model output:
    \begin{equation}  L_3=\ell_1(\textrm{E}(X),\textrm{E}(\tilde{X}))
    \end{equation}
The \textbf{total loss} is given by
    \begin{equation}
       L_{\textrm{Total}} = L_1+L_2+\lambda\cdot L_3
    \end{equation}
which is trained with $\lambda=10000$.

\subsection{Vocoder: The Griffin-Lim Method}
\label{sec:Vocoder}
The output of the decoder is a spectrogram. In order to convert the signal back to the time domain, the phase of the signal is required. As the processed spectrogram is totally different from the input spectrogram, there is no sense in using the original phase, as commonly done in enhancement tasks.

In order to reconstruct the time signal, we used the Griffin-Lim algorithm \cite{6701851}. The gist of this algorithm is to use an initial phase for the signal, whether the original phase, zero-phase, or random phase, and then apply an iterative update until convergence.
An alternative is to use a deep learning-based algorithm, e.g., Hifi-GAN \cite{kong2020hifi}, Mel-GAN \cite{sheng2019high} or other similar solutions \cite{wyse2017audio}. In our tests, we found that the results using the Griffin-Lim algorithm were just as good as the ones using deep-learning methods. We, therefore, preferred this solution due to its relatively low  computational resource requirements.
\section{Datasets}
\label{sec:Datasets}
\subsection{Existing Datasets}
\label{sec:Existing Datasets} 
The corpus used to train our model is \ac{NHSS} \cite{NHSS}, a database of parallel recordings of speech and singing. The audio recordings in the \ac{NHSS} database correspond to a total of 100 songs sung and spoken by 10 singers, 5 male, and 5 female, resulting in a total of 7 hours of audio data. 
Although this dataset is presented as a parallel dataset, we did not use this property in our training but rather  arbitrarily introduced both singing and speaking audio to the model.
 Another corpus used for our model is the \ac{VCTK} Corpus \cite{VCTK}. It includes speech recordings by 109 native English speakers. We use this dataset in inference time as a pool of unseen speakers.
\subsection{New Dataset: `Songlist`}
\label{sec:New Dataset - 'Songlist'}
We created our own dataset and handpicked over 600 songs performed by the top music artists of the past century. We then used ‘Spleeter’ \cite{Spleeter, ca2021research}, which, as explained above, is a tool to separate the song into ‘vocal’ and ‘instrumental’ channels. The vocal channel will be used for the speech-to-singing conversion and later will be re-mixed with the ‘instrumental’ channel.
The songs mainly contain only a single singer, so the model could make a conversion properly between a single speaker and a single singer.

\section{Experimental Study}
\label{sec: Experimental Study}
\subsection{Training Process}
\label{sec: Training Process}
In order to avoid the need for a parallel dataset (speech and singing of the same speaker), the model is trained as an autoencoder model with self-reconstruction loss.
For the train set, we chose to use only the \ac{NHSS} dataset, which combines both speech and singing audio.
During training, we randomly choose an utterance (either speaking or singing) and train only the autoencoder block. This utterance, during training, is simultaneously treated  as both the style and the content. We do not make use of the parallelism in the dataset but rather provide both the speaking and singing examples during train time. This will, hopefully, make the model familiar with both and encourage it to generate a singing-like output at inference time.
It is important to note, that during the training stage, the concatenated embedding vector (to both the encoder and decoder) and the input spectrogram are of the same, randomly chosen, person.

In the test stage of our model, when performing a style transfer, the embedding vector, concatenated to the inputs of the encoder and decoder, is of person B, the speaker, while the input spectrogram is of person A,  the singing person.
We stress that, at inference time, we are not limited by the type or pool of inputs, for both the songs and the speakers. For the results we present here, we used a trained model only over the \ac{NHSS} dataset but took a style and content from the \ac{VCTK}, \ac{NHSS} datasets, and our newly created dataset, the `Songlist'.

\subsection{Results}
\label{sec: Results}
One of the challenges we faced with the evaluation of our results was the fact that we have not seen any published work which attempts to perform the same task, namely applying the style of an unknown speaker to any given song.

We, therefore, resort to a subjective evaluation, based on non-intrusive quality measures.
For the evaluation, we used a group of 25 random listeners, both male and female between the ages of 18-60, who are not experts in music or sound generation. Most of the participants are friends or family, or a second circle to them. During the listening test, the participants listened to two audio clips: first, “Audio A - Content” and second, “Audio B - Style”. Lastly, they listened to the output of our network: “Audio C” an audio clip with the content from “Audio A - Content” in the style of “Audio B - Style”. Then they were asked to answer the following questions:
\begin{enumerate}
    \item From `Audio C' - how clear were the words to you?
    \item From `Audio C' - how much does the melody sound like the melody in `Audio A'?
    \item How much did `Audio C' sound like it was sung in the style of the person in the `Audio A'?
    \item How much did `Audio C' sound like it was sung in the style of the person in `Audio B'?
    \item How life-like did `Audio C' sound (does it sound like a real human being)?
\end{enumerate}
\begin{table}[htbp]
\caption{Listening test survey results. 
$\downarrow$ designate the lower is better, and $\uparrow$ the higher is better.}
\label{tab: Listening test survey results}
 \centering
 \begin{tabular}{c c} 
 \toprule
 Question &Result \\ [1ex] 
 \midrule
 1 $\uparrow$   & 3.55$\pm$0.386 \\ 
 2 $\uparrow$   & 3.87$\pm$0.356 \\
 3 $\downarrow$ & 3.06$\pm$0.448 \\
 4 $\uparrow$   & 3.17$\pm$0.404 \\
 5 $\uparrow$   & 2.39$\pm$0.413 \\ [1ex] 
 \bottomrule
 \end{tabular}
\end{table}
The results are presented in Table~\ref{tab: Listening test survey results}.
We report on the listening test survey, displayed on a 1–5 scale with their respective 95\%  confidence intervals. 

In general, we observe that the main goal was accomplished. We were able to synthesize a result that sounds closer to the style of the target person (3.17) than to the style of the original person (3.06), namely, it sounds more like Person B than Person A. Moreover, the quality of the sound produced was good enough for listeners to clearly hear the words and to perceive that the melody has not changed much throughout the entire process.
That being said, the style change attempted is probably not distinctive enough, especially on a singular test, when the listener is not comparing results to other existing networks, but to their ground truths.

Another aspect that should be improved is the ``livelihood” of the result. What we have seen throughout our work is really how sensitive the human sense of listening is, and how even a good output (both in terms of melody and words), which does sound similar to the target, can still be far from persuading that this is actually a human singing. 
Some samples of the results from our style transfer model can be found online.\footnote{\texttt{https://thesignitproject-singit-streamlit-\\streamlit-code-zgx4p0.streamlit.app/}}
% Old page Aaron create with github
%\footnote{\texttt{https://anonymous.4open.science/w/SingIt/}}
%
% New page with streamlit:

\section{Summary}
In this paper, we have presented a new way of generating singing from one's voice while only being familiar with the user's speech. Current solutions deal with the complexity of this problem by requiring a singing utterance from the user. Hence they are only addressing the problem of generating a singing voice from another singing voice or the problem of generating only specific singers or songs that have been learned by the network. 
Our proposed method involves separating the vocals, then using an encoder and a decoder in the frequency domain and reconstructing the signal using a Griffin-Lim decoder. 
The results presented show how we managed to transfer the style from a given speech and apply it to a song. Our results show alignment with the original song both in pitch and tempo. These results still have much to improve in the ``livelihood" aspect of the output audio, as well as the capability to succeed with the style transfer in more complex cases, e.g., more singers, and inferior quality input.
The implementation of this work can be relevant in a wide spectrum of fields from art to science, to health and pleasure, and will hopefully invite much more work to be done in this specific field.
 
 \balance
\bibliographystyle{IEEEtran}
\bibliography{my_library}

\end{document}

%% file: acronyms.tex
\acrodef{NPSS}{Neural Parametric Singing Synthesizer}
\acrodef{NHSS}{NUS-HLT Speak-Sing}
\acrodef{VCTK}{Voice Cloning Toolkit}

\acrodef{BLSTM}{Bidirectional Long Short-Term Memory}
\acrodef{CNN}{Convolutional Neural Networks}

\acrodef{STFT}{Short-time Fourier Transform}
\acrodef{TF}{Time-Frequency}